\documentclass[prb,preprint,showpacs,amsmath,amssymb,superscriptaddress,longbibliography]{revtex4-1}
\usepackage{graphicx}
\usepackage{breqn}
\usepackage{xcolor}
\usepackage{color}
\usepackage{booktabs}
\usepackage{float}
\usepackage{longtable}
\usepackage{epsfig}
\usepackage{bm}
\usepackage{dcolumn}
\usepackage{lipsum, babel}
\usepackage{soul}

\usepackage{rotating} % for sidewaystable

\usepackage[normalem]{ulem}
\usepackage{epstopdf}
\makeatletter
\let\cat@comma@active\@empty
\makeatother
\begin{document}

\title{Altermagnetic spintronics}

\author{T.~Jungwirth}
\thanks{email: jungw@fzu.cz, ORCID: 0000-0002-9910-1674}
\affiliation{Institute of Physics, Czech Academy of Sciences, Cukrovarnick\'a 10, 162 00, Praha 6, Czech Republic}
\affiliation{School of Physics and Astronomy, University of Nottingham, Nottingham NG7 2RD, United Kingdom}
\affiliation{Center for Science and Innovation in Spintronics, Tohoku University, 2-1-1 Katahira, Aoba-ku, Sendai 980-8577, Japan}

\author{J.~Sinova}
\affiliation{Institut für Physik, Johannes Gutenberg Universität Mainz, D-55099 Mainz, Germany}
\affiliation{Department of Physics, Texas A \& M University, College Station, Texas 77843-4242, USA}

\author{P.~Wadley}
\affiliation{School of Physics and Astronomy, University of Nottingham, Nottingham NG7 2RD, United Kingdom}

\author{D.~Kriegner}
\affiliation{Institute of Physics, Czech Academy of Sciences, Cukrovarnick\'a 10, 162 00, Praha 6, Czech Republic}

\author{H.~Reichlová}
\affiliation{Institute of Physics, Czech Academy of Sciences, Cukrovarnick\'a 10, 162 00, Praha 6, Czech Republic}

\author{F.~Krizek}
\affiliation{Institute of Physics, Czech Academy of Sciences, Cukrovarnick\'a 10, 162 00, Praha 6, Czech Republic}

\author{H.~Ohno}
\affiliation{Center for Science and Innovation in Spintronics, Tohoku University, 2-1-1 Katahira, Aoba-ku, Sendai 980-8577, Japan}
\affiliation{Laboratory for Nanoelectronics and Spintronics, Research Institute of Electrical Communication, Tohoku University, 2-1-1 Katahira, Aoba-ku, Sendai 980-8577, Japan}
\affiliation{Advanced Institute for Materials Research, Tohoku University, 2-1-1 Katahira, Aoba-ku, Sendai 980-8577, Japan}
\affiliation{Center for Innovative Integrated Electronic Systems, Tohoku University, 468-1 Aoba, Aramaki, Aoba-ku, Sendai 980-8572, Japan}

\author{L.~Šmejkal}
\affiliation{Max Planck Institute for the Physics of Complex Systems, N\"othnitzer Str. 38, 01187 Dresden, Germany}
\affiliation{Max Planck Institute for Chemical Physics of Solids, N\"othnitzer Str. 40, 01187 Dresden, Germany}
\affiliation{Institute of Physics, Czech Academy of Sciences, Cukrovarnick\'a 10, 162 00, Praha 6, Czech Republic}

%

%\date{\today}

\begin{abstract}

A recently discovered form of magnetism, known as altermagnetism, is characterized by an unconventional $d$-wave (or higher even-parity-wave) collinear compensated spin ordering. This phase supports strongly spin‑polarized currents in the absence of net magnetization and exhibits fast spin dynamics. Meanwhile, spintronic memories based on conventional ferromagnets are transitioning from a niche technology to mass production of components for advanced‑node microprocessor chips. In this perspective, we outline how the distinct signatures of altermagnetism may broaden the scope of spintronics research and influence the functionality and scalability of future devices. In particular, we focus on the emerging theoretical concepts in altermagnetic spintronics and on the potential interplay with ferroelectricity or superconductivity. We also provide an outlook on the growing experimental research on altermagnetic spintronics, and on the role of relativistic spin-orbit coupling phenomena.

\end{abstract}

\maketitle

%\subsection*{Introduction}

Altermagnetism has recently attracted broad attention because it enriches the landscape of conventional ferromagnetic and antiferromagnetic phases by a distinct unconventional symmetry class \cite{Smejkal2021a,Smejkal2022a}. It combines strong spin polarization, typical of ferromagnets, with vanishing net magnetization, characteristic of antiferromagnets, and with additional features unparalleled in the conventional magnetic phases. This opens prospects of research directions and potential technological applications beyond the realm of conventional magnets\cite{Smejkal2022a,Smejkal2022AHEReview,Bai2024,Song2025,Jungwirth2025,Jungwirth2025a,Cho2024}. 

Altermagnetism is rigorously classified and described by spin symmetries, which establish that, apart from the conventional ferromagnetism and antiferromagnetism, there exists a third distinct collinear magnetic phase\cite{Smejkal2021a,Smejkal2022a}. The ordering in the conventional ferromagnets (Fig.~1a), leads to a spin polarized electronic structure with a net magnetization, breaking time-reversal ($\cal{T}$) symmetry. The conventional antiferromagnetic ordering generates electronic bands which  remain spin unpolarized, resulting in a zero net magnetization (Fig.~1b). The altermagnetic ordering\cite{Smejkal2021a,Smejkal2022a} is unconventional in the sense that its spin symmetries give zero net magnetization but at the same time allow a spin-polarized electronic structure, which has anisotropic $d$-wave ($g$ or $i$-wave) symmetry and breaks the $\cal{T}$-symmetry\cite{Smejkal2020,Mazin2021,Smejkal2022AHEReview,Smejkal2021a,Smejkal2022a,Jungwirth2025,Jungwirth2025a}. The $d$-wave altermagnetic symmetry is characterized by 2 (4 or 6 for $g$ or $i$-wave, resp.)  spin-degenerate nodal surfaces in the momentum-space Brillouin zone, and a strong non-relativistic spin polarization of electronic bands whose sign alternates from one to the  other side of each nodal surface (Fig.~1c).

 In general,  altermagnetism  has been predicted in  materials including three-dimensional \cite{Smejkal2021a,Smejkal2022a,Guo2023b,Xiao2024,Bai2024} and two-dimesional\cite{Smejkal2022a,Smejkal2022GMR,Ma2021,Egorov2021,Brekke2023,Cui2023,Chen2023b,Mazin2023a,Sodequist2024} inorganic, as well as organic\cite{Naka2019,Ferrari2024} crystals. Altermagnetic conduction types can range from insulators and semiconductors to metals and superconductors, and altermagnetism can occur in crystals with common light elements and strong spin-coherence\cite{Smejkal2022a}. 

Experimentally, the unconventional $\cal{T}$-symmetry breaking spin polarization, generated by a collinear compensated magnetic order, was confirmed by angle-resolved photoemission spectroscopy  measurements in semiconducting and metallic room-temperature altermagnets MnTe and CrSb \cite{Krempasky2024,Lee2024,Osumi2024,Reimers2024,Yang2024,Ding2024,Zeng2024}. Spectroscopic signatures of altermagnetism have been also reported in metallic KV$_2$Se$_2$O and RbV$_2$Te$_2$O \cite{Jiang2025,Zhang2025a} whose monolayers have analogous crystal and magnetic structure to the 2D model in Fig.~1c. 

 The above characteristics of altermagnetism, together with their THz-range spin dynamics\cite{Smejkal2023,Liu2024b}, as typical in compensated magnets, are favorable for the research towards scalable spintronic technologies\cite{Smejkal2022a}.  Spintronics based on conventional ferromagnets is currently transitioning from a niche to a mass-production integrated circuit technology.  This is due to  the magnetic random-access memories (MRAMs), which represent a scalable, low-cost, leakage-free, power-efficient and radiation-hard technology for non-volatile embedded memories  on microprocessor  chips, where they start to replace embedded Flash memories\cite{Worledge2022}. Simultaneously, big data and artificial intelligence are boosting demand for high-capacity, fast and energy-efficient memory technologies. Whether altermagnets can help future spintronics to meet these challenges is one of the questions motivating the research of this class of magnets.

In the following sections we first briefly summarize, as a reference, the present ferromagnetic spintronic technology,  and proof-of-concept spintronic devices  based on conventional antiferromagnets  and non-collinear compensated magnets.  We then describe  the proposed basic physical concepts of the altermagnetic spintronics, and outline their possible extensions exploiting the compatibility of altermagnetism with ferroelectricity or superconductivity. We conclude with a brief discussion of initial experiments and  the role of relativistic and topological effects in altermagnetic spintronics. 
 
%----------------------------

\subsection*{Ferromagnetic spintronics}

 The spintronic MRAM represents a revolutionary magnetic information technology (IT). In comparison, the established role of magnets in modern IT is in data storage on ferromagnetic hard-disk drives. Similar to magnetic-relay bits of the early computing machines,  hard-disk drives employ magnetic fields and mechanically moving parts for their operation. The spintronic MRAM does not use magnetic fields and is implemented on a solid-state integrated circuit.  Instead of magnetic fields,  MRAMs use the strong spin polarization of the electrical current, provided by the ferromagnetic order, for both reading and writing. In an MRAM bit, passing the spin-polarized current between reference and recording electrodes, separated by a tunnel barrier, senses whether the magnetizations of the two ferromagnetic electrodes are anti-parallel (high resistance) or parallel (low resistance). The corresponding tunneling magnetoresistance (TMR) readout signal, defined as a relative difference between the high and low-resistance states, can reach hundreds of per cent\cite{Bhatti2017}. At a larger amplitude of the applied current, writing is facilitated by the spin-transfer torque (STT) effect, which switches the magnetization of the recording electrode back and forth, depending on the direction of the writing current\cite{Bhatti2017}. 

An alternative three-terminal MRAM geometry is based on writing by the relativistic spin-orbit torque (SOT)\cite{Prenat2016,Bhatti2017}. Here, the ferromagnetic electrode is switched by sending an electrical current along the plane of the device. This electrical current generates a spin current -- due to the spin‑Hall effect\cite{Sinova2015}  -- in a nearby spin-orbit coupled material, and that spin current triggers the switching of the magnet. Unlike the compact two-terminal structure of the STT-MRAM bit, the SOT-MRAM bit is a three-terminal devices and, thus, has a larger bit-cell size.

The sensitivity of ferromagnets to magnetic fields and the stray fields generated by ferromagnets, which enable the hard-disk  magnetic-storage technology, are thus not needed for either reading or writing in the spintronic MRAMs. They remain merely a limiting scalability factor which is suppressed by elaborate multilayer-stack designs of the present out-of-plane magnetized MRAM bits.

Ferromagnets in principle enable to approach the fundamental physical limit of the minimum heat generation when operating a non-volatile bit.  According to the Landauer's thermodynamic principle\cite{Landauer1961}, this minimum energy dissipation is given by $k_B T \ln 2$, where $T$ is the temperature and $k_B$ is the Boltzmann constant. At room temperature, the Landauer limit is  in the meV range. On one hand, the operation at the Landauer  limit was so far demonstrated\cite{Hong2016}, using a quasi-equilibrium switching protocol\cite{Landauer1961}, in exploratory ferromagnetic bits at switching times above $\mu$s, which implies switching speeds below MHz and thus orders of magnitude below the GHz speed of commercial MRAM chips. On the other hand, the Joule heating, $\sim R I^2 t_p$, in MRAM bits is many orders of magnitude above the Landauer limit\cite{Bhatti2017}. Here $R$ is the resistance of the bit, $t_p$ is the writing pulse-time, and $I$ is the required writing current whose amplitude is nearly independent of $t_p$ as long as the pulse-time is comparable or larger than the coherent spin-precession time-scale\cite{Bedau2010,Prenat2016,Garello2014}. In ferromagnets, this threshold time-scale is typically in the ns-range, corresponding to the GHz-range of the ferromagnetic resonance frequency governed by the weak relativistic magnetic-anisotropy energy. Instead of continuing to decrease linearly with shorter $t_p$, the Joule heating increases by three orders of magnitude when moving from ns to ps writing pulse-times. This happens because, below the ns threshold, the required amplitude of the writing current becomes inversely proportional to the pulse duration\cite{Prenat2016,Garello2014}.

The quantum uncertainty principle imposes an additional fundamental constraint on the relationship between the minimum operation time and generated heat. A writing single-photon pulse in the form of a narrow wave-packet peaked at the meV energy of the Landauer limit, i.e. peaked at the THz frequency, can have its pulse-length scaled down at most to the picosecond range. The simultaneously least-dissipative and fastest control of a bistable memory bit is thus in the picosecond (THz) range\cite{Landauer1961,Schlauderer2019,Kimel2020}, which is beyond the reach of the present  ferromagnetic MRAMs. This motivates research of spintronics beyond ferromagnets.

\subsection*{Spintronics with conventional antiferromagnets and non-collinear compensated magnets} 

Compensated magnets, such as the conventional collinear antiferromagnets, can share with ferromagnets the non-volatile bi-stability. In contrast to ferromagnets, however, the resonance frequencies are in the THz range due to the strong inter-spin-sublattice exchange energy.  Moreover, compensated magnets  with their vanishing net magnetization generate negligible stray fields and are insensitive to perturbing magnetic fields. Over the past two decades, these favorable features have motivated  research in spintronic devices whose functionality is governed by  compensated magnets.

The field gained particular momentum after the experimental demonstration of electrical writing and reading at room temperature in memory bits fabricated from a collinear antiferromagnet CuMnAs\cite{Wadley2016} on chips that can be  controlled from a PC via a standard USB interface\cite{Olejnik2017}. Subsequent THz experiments\cite{Olejnik2018} showed that the switching pulse-length can be reduced to the ps-scale and, by using THz photoconductive switches, the ultra-fast switching was simultaneously spatially controlled down to the nano-scale\cite{Heitz2021}. Unlike ferromagnets, the dissipated energy did not show the unfavorable increase for the pulse length reduced from nanoseconds to picoseconds. However, it remained on the scale of 10~meV per atom, which is still far from the Landauer limit of meV per bit. 

In the experimental memory bits realized in CuMnAs or other conventional collinear antiferromagnets, reading and writing is typically realized via $\cal{T}$-symmetric responses. These can include a combination of anisotropic magnetoresistance and SOT, or ordinary resistivity changes induced by  switching into magnetic nano-textured multiple-metastable states\cite{Wadley2016,Kaspar2021}. The weak relativistic origin of the responses in the former case, and heating close to the N\'eel temperature  in the latter case limit  efficiency. Using these relatively inefficient  $\cal{T}$-symmetric responses is dictated by the nature of the magnetic ordering in the conventional collinear antiferromagnets, with opposite spins in the crystal related by either translation or inversion. Such ordering does not generate the strong $\cal{T}$-symmetry breaking and spin polarization in the non-relativistic electronic spectra, and the corresponding efficient  $\cal{T}$-symmetry breaking responses, typical of ferromagnets. Only weak $\cal{T}$-symmetry breaking transport effects were detected in collinear antiferromagnets,  because in these magnets they require a combination of the relativistic spin-orbit coupling and a non-linear response to the applied electric bias\cite{Godinho2018}.  A broader overview of antiferromagnetic spintronics -- including ultrafast dynamics and switching -- can be found in Refs.~\onlinecite{Jungwirth2016,Zelezny2018,Nemec2018,Kimel2020}.

Electrical readout by $\cal{T}$-symmetry breaking linear responses, including the relativistic anomalous Hall effect (AHE) or the non-relativistic TMR, was demonstrated in  the family of Mn$_3$X (X=Sn,Pt,...) non-collinear compensated magnets, in combination with electrical switching\cite{Tsai2020,Takeuchi2021,Pal2022,Chen2023,Qin2023}. Despite the vanishing net magnetization, the compensated non-collinear magnetic order can generate $\cal{T}$-symmetry breaking electronic band structure and  spin-dependent transport phenomena, reminiscent of ferromagnets\cite{Shindou2001,Chen2014,Kubler2014,Nakatsuji2015,Zelezny2017a,Kimata2019a}. However,   due to the non-collinearity of the magnetic order, spin-up and spin-down electronic states are strongly mixed, which is reminiscent  of the relativistic spin-orbit coupling. This contrasts with the collinear magnetic order in ferromagnets, 
which generates well separated and conserved
spin-up and spin-down transport channels underpinning   the strong TMR and
STT responses for reading and writing in MRAMs. An overview of physics and devices based on the non-collinear compensated magnets can be found in Refs.~\onlinecite{Smejkal2022AHEReview,Rimmler2024,Han2025}.)

\subsection*{Basic concepts of altermagnetic spintronics}

Altermagnets enable a favorable combination of THz-range resonance frequency, lack of magnetic stray fields, and  $\cal{T}$-symmetry breaking linear responses, with a collinear magnetic order  and the corresponding well separated and conserved spin-up and spin-down transport channels. Figure~1c shows position-space and momentum-space schematics of the $\cal{T}$-symmetry breaking spin-dependent currents in a model $d$-wave altermagnet. For the electric bias $E$ applied along the $x$ ($k_x$) axis, the linear-response electrical current is dominated by electrons form the spin-up crystal sublattice (spin-up Fermi surface). Following the $d$-wave symmetry of the altermagnetic order, in which opposite-spin electronic states are related by a crystal-rotation symmetry, the spin polarization of the electrical current reverses from up to down when the electric bias is rotated to the  $y$ ($k_y$) direction\cite{Gonzalez-Hernandez2021}. In contrast, the model ferromagnet in Fig.~1a has only one (spin-up) magnetic atom in the unit cell, which determines the (spin-up) polarization of the electrical current, independent of the direction of the applied bias. In the model of a conventional antiferromagnet in Fig.~1b, the translation symmetry relating the opposite-spin sublattices implies that both sublattices contribute equally to the electrical current, rendering the current spin unpolarized for any direction of the applied bias.

In the model $d$-wave altermagnet, the longitudinal electrical current for $E$ applied along the in-plane diagonal direction is spin unpolarized, reminiscent of the conventional antiferromagnet. However, the electrical currents of the spin-up and spin-down electrons are deflected by opposite angles from the diagonal direction of the applied bias\cite{Gonzalez-Hernandez2021}, resulting in a transverse pure spin current\cite{Gonzalez-Hernandez2021,Naka2019}. This non-relativistic phenomenon, dubbed spin-splitter effect (SSE), is reminiscent of the relativistic SHE\cite{Sinova2015}. However the transverse spin currents in the two cases have a principally distinct phenomenology\cite{Gonzalez-Hernandez2021}.  The SSE breaks the $\cal{T}$-symmetry and the polarization axis of the spin current is determined by the spin-polarization axis of the altermagnetic order. In contrast, the SHE is  a $\cal{T}$-symmetric response with the polarization of the spin current determined by the direction of the applied electric bias. In terms of the effects' amplitudes, the SSE is predicted to outperform the SHE by two orders of magnitude because of its strong non-relativistic magnetic-ordering origin\cite{Gonzalez-Hernandez2021}. Finally, for $E$ applied along the out-of-plane direction of the model altermagnet in Fig.~1c, both spin-up and spin-down electrical currents  propagate along $E$, resulting in no transverse spin current and an unpolarized longitudinal current.

The spin-polarized electrical currents were predicted to enable the giant magnetoresistance (GMR) effect in a stack comprising two altermagnetic layers separated by a metallic non-magnetic spacer\cite{Smejkal2022GMR}. In analogy to ferromagnets, the altermagnetic GMR in the current-in-plane  geometry, illustrated in Fig.~2a,  is determined by the ratio of the spin-up and spin-down electrical conductivities for the given direction of the applied bias\cite{Smejkal2022GMR}. 

When the altermagnetic order is orientated such that a spin-polarized electrical current is allowed in the out-of-plane direction, as shown in Fig.~2b,  altermagnetism also enables the counterparts of the ferromagnetic current-perpendicular-to-plane (CPP) GMR for the metallic spacer, and the TMR for an insulating spacer\cite{Smejkal2022GMR}. Remarkably, altermagnetic TMR effects of comparable  amplitudes to ferromagnets are also predicted in the geometry where the total out-of-plane current is spin unpolarized\cite{Smejkal2022GMR,Shao2021}. This is because even in this so called spin-neutral-current geometry, there are transport channels at individual momenta which are spin polarized, as illustrated in Fig.~2c. We also emphasize that, while the GMR determined by the spin-dependent momentum-integrated conductivities is symmetry-restricted to $d$-wave altermagnets, the TMR is allowed by symmetry for any crystal orientation and in all $d$, $g$ or $i$-wave altermagnets\cite{Smejkal2021a,Smejkal2022GMR}.

Besides the above analogies and distinctions from spintronics based on metallic ferromagnets, altermagnetism enables to readily extend the concepts to insulators. Similar to  conventional antiferromagnets, and unlike ferromagnets, the altermagnetic ordering is common in insulators \cite{Smejkal2021a,Smejkal2022a}. In Fig.~3 we show, next to the stacks with metallic altermagnets (Fig.~3a), examples of TMR multilayers with insulating altermagnets (Fig.~3a). Here an unpolarized electrical current injected from a non-magnetic metallic electrode is spin-filtered through the thin-film altermagnetic insulator. The spin-filtering effect depends on whether the magnetic orders in the two altermagnetic layers are arranged parallel or antiparallel, resulting in the TMR effect\cite{Samanta2025}.

Besides the GMR/TMR read-out of the altermagnetic state, the spin-polarized longitudinal and transverse electrical currents in altermagnets are predicted to provide also efficient means for switching the magnetic order or for controlling altermagnetic textures by current-induced spin-transfer torques\cite{Gonzalez-Hernandez2021,Vakili2025,Zarzuela2025}. 

\subsection*{Interplay of altermagnetism with ferroelectricity}

Insulating altermagnets open a possibility to eliminate the parasitic energy dissipation associated with Joule heating. Instead of the dissipative electrical currents, the altermagnetic insulators can be switched by electric fields. A particularly efficient and non-volatile switching mechanism is predicted in insulating multiferroic materials with coexisting altermagnetic and  ferrolelectric orders (Fig.~3b)\cite{Gu2025,Smejkal2024}. Note here that ferromagnetic ferroelectrics are elusive, again because of the general incompatibility of the ferromagnetic ordering with insulating electronic structures\cite{Ramesh2007}. 

If the altermagnetic electrode in a TMR stack is multiferroic, it can be switched by an electric field instead of current via the so-called altermagneto-electric effect\cite{Smejkal2024},  illustrated in Fig.~3c. To explain the phenomenon, we first highlight that the local atomic spin density in an altermagnet can be decomposed into an isotropic dipole contribution and an anisotropic higher-partial-wave component ($d$-wave component for the model $d$-wave altermagnet in Fig.~3c)\cite{Smejkal2020,Jaeschke-Ubiergo2025}. While the dipoles are ordered antiparallel between the neighboring magnetic atoms, i.e. have the Néel-like order, the higher-partial-wave components are identical, i.e., are ordered ferroically. This latter non-Néel ferroic component is one of the key signatures of the altermagnetic ordering\cite{Smejkal2020,Jaeschke-Ubiergo2025}. Its sign is determined in the model altermagnet by the sense of the uniaxial anisotropic distortion of the local crystal environment of the magnetic atoms.  When the altermagnet is ferroelectric, and the sense of the uniaxial distortion reverses between the two magnetic atoms upon reversing the electric polarization $P$, the sign of the higher-partial-wave component of the local spin density flips (Fig.~3c). Because the sign of the higher-partial-wave spin-density component in the position space determines the sign of the alternating spin polarization of the energy iso-surfaces in the momentum space, the latter also flips when reversing $P$. 

In the context of the insulating altermagnets, we next recall a recent study of TmFeO$_3$ \cite{Schlauderer2019}. The experiment in this altermagnet demonstrated a reorientation of the axis of the ordered spins by a ps-long THz-field pulse at energy   of $\sim$meV per atom. The study was performed at low temperatures, and only a transient reorientation was demonstrated without accomplishing a stable non-volatile switching. Nevertheless, it showed that reaching the  Landauer limit is becoming a legitimate, albeit still challenging  goal in the research of ultra-fast altermagnetic spintronics.

\subsection*{Interplay of altermagnetism with superconductivity}

The predicted altermagnetic order in the insulating parent cuprate La$_2$CuO$_4$ of high-T$_c$ superconductors pointed towards yet another research avenue, this time exploring and exploiting the interplay of altermagnetism with superconductivity\cite{Smejkal2021a,Smejkal2022a}. The research direction offers alternative routes to suppress Joule heating, to generate strong spin-dependent  responses and, in a broad sense, to merge the fields of spintronics and superconducting quantum technologies. 

An example is the  infinite magnetoresistance (IMR) effect, predicted in a stack comprising two metallic altermagnetic layers separated by a spacer made of a conventional $s$-wave superconductor (Figs.~2a and 3a)\cite{Giil2024}. Here the altermagnet suppresses the superconducting transition temperature via the inverse proximity effect. The strength of the suppression depends on whether the mutual magnetic configuration of the two altermagnetic electrodes is parallel or antiparallel, rendering the spacer superconducting in one configuration and, at the same temperature, non-superconducting  in the other configuration.  The IMR could enable a large readout signal alternative to GMR for cryogenic stray-field-free memory devices.

In the IMR effect, the altermagnetic order plays merely a role of suppressing the $s$-wave order parameter in the adjacent conventional superconductor.  Other proposed device schemes employ the unconventional $d$-wave (or higher even-parity-wave) altermagnetic symmetry  which can support unconventional spin-triplet or finite-momentum Cooper pairing (cf. Figs.~4b,d and Fig.~4a). The superconducting spin-splitter effect\cite{Giil2024a}, schematically shown in the left column of Fig.~4c, is an example of such device concepts. The phenomenon is based on the predicted unconventional spin-triplet Cooper-pairing correlations enabled by the spin-split altermagnetic electronic structure. As a result, an applied supercurrent from a conventional superconducting electrode adjacent to the altermagnet in the geometry analogous to the dissipative SSE induces an edge magnetization with opposite signs at opposite edges of the altermagnet. Also in analogy to the dissipative SSE, the sign of the edge magnetization  reverses when reversing the supercurrent. 

In  Figs.~4e,f, we show two examples of superconducting quantum phenomena based on unconventional spin-singlet Cooper pairing with a finite center-of-mass momentum (Fig.~4d), which can be also  promoted by the altermagnetic symmetry of the spin-split Fermi surfaces. A superconducting  order parameter corresponding to the finite-momentum Cooper pairing is spatially inhomogeneous and contains nodes at which the order parameter changes the phase by $\pi$. As a result, a Josephson junction  comprising two conventional superconducting electrodes separated by an altermagnet shows $0-\pi$ oscillations as a function of the length of the altermagnet\cite{Ouassou2023,Zhang2023,Beenakker2023}. Here the Josephson-junction ground state oscillates between having the same phase of the order parameters in the two superconducting electrodes, or having a phase difference of $\pi$. Compared to ferromagnetic Josephson junctions, with a typically exponential decay superimposed on the $0-\pi$ oscillations, the altermagnetic Josephson junctions can show undamped $0-\pi$ oscillation over a large range of Josephson-junction lengths which is favorable for their experimental realization\cite{Ouassou2023}. 

In altermagnets,  the center-of-mass momentum of the Cooper pair locks to discrete crystal axes, which eliminates the rotational Goldstone mode and preserves a sizable superconducting gap\cite{Yang2025}. Moreover, besides being proximity induced, it was also predicted that the unconventional spin-singlet finite-momentum superconductivity, namely the Fulde-Ferrell-Larkin-Ovchinnikov phase, can coexist with altermagnetism in the same crystal\cite{Sumita2023,Chakraborty2025,Sim2025}. Both the proximity-induced and the intrinsic superconductivity cases are predicted to lead to non-reciprocal transport due to the broken time-reversal and inversion symmetries by the combined order parameters\cite{Banerjee2024a,Chakraborty2025,Yang2025,Sim2025}. In the corresponding  superconducting diode effect, a dissipationless supercurrent flows in one direction, while  a dissipative normal current flows in the opposite direction (Fig.~4f).  In presence of an external magnetic field,  the superconducting diode efficiency in altermagnets can, in principle, reach 100\%, corresponding to the superconducting critical current going to zero for one of the two opposite current directions\cite{Chakraborty2025}. 

The altermagnetic $0-\pi$ Josephson junctions and superconducting diodes have attracted interest as testbeds of the proximity-induced or intrinsic unconventional Cooper pairing in altermagnets. Simultaneously,  they are attractive for their potential applications in cryogenic low-dissipation electronics and spintronics, and quantum sensing and computing technologies. Additional tuneability of these devices can be achieved through gating or doping when employing altermagnetic semiconductors. Because altermagnetism is readily compatible with semiconducting band structures, it circumvents the limitations of ferromagnetic semiconductors,  in which enhancing the Curie temperature towards room temperature tends to  diminish the semiconducting properties\cite{Dietl2014,Jungwirth2014}.  MnTe is a prototypical room-temperature altermagnetic semiconductor whose intrinsic band structure features a band-gap and an alternating ($g$-wave) spin splitting, both on the favorable $\sim 0.1-1$eV scale\cite{Smejkal2021a}.

\subsection*{Discussion of initial experiments and relativistic effects}

In the previous section, we focused on phenomena and device functionalities based on strong non-relativistic physics of the altermagnetic ordering. This strong altermagnetic ordering originates from the interplay of the single-particle crystal potential of a desired symmetry and many-body electron-electron interactions\cite{Smejkal2021a,Smejkal2022a,Jungwirth2025,Jungwirth2025a}. The crystal-potential and the electron-electron interaction terms in the non-relativistic Hamiltonian preserve the spin-space rotation and crystal rotation symmetries, and the $\cal{T}$-symmetry. The altermagnetic ground-state breaks these symmetries spontaneously. In contrast, the single-particle spin-orbit coupling term in  the $1/c^2$ expansion of the Dirac equation breaks the spin-space rotation symmetry already on the level of the (relativistic) Hamiltonian. Similarly, the single-particle Zeeman term in the Hamiltonian, describing the dipolar coupling of the spin and the magnetic field, breaks the spin-space rotation symmetry, as well as the $\cal{T}$-symmetry. Experimentally, it is thus important to distinguish between effects caused by the spontaneous symmetry breaking due to the altermagnetic  ordering,  and those arising from symmetry breaking due to the relativistic terms in the Hamiltonian. However, the altermagnetic ordering can also interact constructively with relativistic and topological effects, giving rise to further spintronic functionalities.

The disentanglement of the non-relativistic and relativistic effects have been of concern since initial experiments exploring the predicted spin-dependent transport phenomena in altermagnets \cite{Smejkal2020,Smejkal2021a,Gonzalez-Hernandez2021,Bose2022,Bai2022,Karube2022,Feng2022}. In measurements searching for the SSE in the  theoretically proposed $d$-wave altermagnetic candidate RuO$_2$\cite{Smejkal2020,Smejkal2021a,Gonzalez-Hernandez2021,Bose2022,Bai2022,Karube2022}, the competing SHE induced by the relativistic spin-orbit coupling was found to play a major role\cite{Wang2024}.  

Similar care has to be taken in measurements of the AHE in presence of an external magnetic field due to the competing contribution from the Zeeman term, as was the case of the initial AHE measurements performed  in  RuO$_2$\cite{Feng2022}. Studies of this altermagnetic candidate are further complicated by an ongoing debate on the intrinsic vs. extrinsic origins of the measured magnetic signals in some of the investigated samples, or by the absence of any detected magnetic response in others\cite{Berlijn2017a,Zhu2018,Feng2022,Bose2022,Bai2022,Karube2022,Lovesey2023c,Liu2023,Fedchenko2024,Smolyanyuk2023,Kessler2024,Fan2024,Wenzel2025,Jeong2025a,Hiraishi2024,Wang2024,Noh2025}.  The seeming controversy in the reports has been attributed to a fragile magnetic ordering  in the correlated ruthenates, including RuO$_2$, which may be affected by  the film-thickness, strain, impurities, or disorder\cite{Smolyanyuk2023}.  

Zero-field remanent AHE signals, excluding the competing Zeeman effect, were detected in an altermagnetic candidate thin-film Mn$_5$Si$_3$\cite{Reichlova2024}, and altermagnetic MnTe, VNb$_3$S$_6$ or FeS whose robust collinear compensated magnetic orders were confirmed by neutron diffraction measurements\cite{Betancourt2021,Kluczyk2024,Ray2025,Takagi2025}. The AHE in collinear magnets, including altermagnets, vanishes by symmetry in the absence of the relativistic spin-orbit coupling\cite{Smejkal2022AHEReview}. This makes the AHE an example of a constructive interplay of the $\cal{T}$-symmetry breaking by the altermagnetic ordering and the relativistic spin-orbit coupling. 

We also point out that the Hall vector requires the same broken symmetries as the magnetization vector. In other words, when the relativistic AHE is allowed by symmetry, the weak relativistic magnetization is also allowed in the altermagnet. However, the remanent  AHE in altermagnets is not generated by $\cal{T}$-symmetry breaking due to the weak magnetization, but rather by the strong $\cal{T}$-symmetry breaking induced by the compensated altermagnetic ordering, as shown theoretically for the intrinsic Berry-curvature mechanism of the AHE in a number of altermagnets\cite{Smejkal2020,Mazin2021,Reichlova2024,Betancourt2021,Smejkal2020,Smejkal2022AHEReview,Takagi2025}. 

Unlike the dc AHE, the ac $\cal{T}$-symmetry breaking linear responses, such as the  X-ray magnetic circular dichroism (XMCD), enable to disentangle the contributions due to the altermagnetic ordering and the weak magnetization by their distinct spectral dependences. This was theoretically and experimentally demonstrated in MnTe\cite{Hariki2023}, showing that the altermagnetic contribution dominates in MnTe at zero field, while  the magnetization contribution takes over at applied fields of a few tesla. 

In combination with the photoemission electron microscopy, the X-ray magnetic circular dichroism was used to monitor the controlled preparation of single-domain altermagnetic states in MnTe, as well as predesigned domain walls and topological vortices\cite{Amin2024}. The demonstrated control of the altermagnetic domains and textures is a key prerequisite for the future systematic experimental exploration of the proposed altermagnetic-spintronic device concepts.

Another example of the constructive interplay of the altermagnetic ordering and the spin-orbit coupling is the spin-dependent conductivity and the related SSE in $g$ and $i$-wave altermagnets, where these effects vanish by symmetry in the non-relativistic limit. In the $g$-wave altermagnet MnTe with the magnetic easy-axis in the $c$-plane of the hexagonal crystal, the relativistic conductivity tensor in the $c$-plane is spin dependent for the spin component along the $c$-axis\cite{Betancourt2021}. This  relativistic spin-polarization orthogonal to the in-plane magnetic easy-axis  can be understood by exploring the effect of the spin-orbit coupling in the momentum-space nodal plane parallel to the $c$-plane of this $g$-wave altermagnet. The spin-orbit coupling lifts the spin degeneracy in this nodal plane\cite{Krempasky2024}, and generates a $d$-wave like alternating spin polarization along the $c$-axis, enabling the corresponding spin-dependent conductivity and the SSE. The collinear relativistic spin-polarization in the nodal plane, in other words the absence of a spin texture with a momentum dependent spin-axis, is symmetry protected in MnTe. Simultaneously, the amplitude of the relativistic spin splitting is large, reaching $\sim 100$~meV due to to the heavy Te atoms. The collinearity and the large spin-splitting amplitude strengthen the resulting relativistic spin-dependent transport effects.  

Until the discovery of altermagnetism, the experimental realization of strong and collinear spin polarization induced by the spin-orbit coupling had been  long-sought  goal for the research of relativistic spintronics. Besides the diffusive spin-dependent conductivity mentioned above, the strong collinear relativistic spin polarization opens the possibility in altermagnetic 2D topological insulators to realize  robust non-dissipative spin-transport phenomena, including the quantum spin (or anomalous) Hall effect\cite{Mazin2023a}.

The altermagnetic phenomena and device concepts outlined in this perspective are likely to draw wide experimental interest and may drive advances in scalable spintronics as well as in superconducting and topological quantum technologies.

\subsection*{Author contributions}
T.J. wrote the initial draft based on inputs from all co-authors.  All authors contributed to the selection of the content, and read and commented on the paper.

\subsection*{Competing interests }
The authors declare no competing interests.

\subsection*{Funding statement}
T.J. discloses support for this work from the Ministry of Education of the Czech Republic Grant No. CZ.02.01.01/00/22008/0004594 and from the European Research Council Advanced Grant No. 101095925. J.S. and L.Š. disclose support for this work from Deutsche Forschungsgemeinschaft (DFG, German Research Foundation) - DFG Project 452301518, TRR 288 – 422213477 (project A09) and TRR 173 – 268565370 (project A03). L.Š. discloses support for this work from the European Research Council Starting Grant No. 101165122. D.K. discloses  support for this work from the Czech Science Foundation Grant No. 22- 22000M, and from the Czech Academy of Sciences Lumina Quaeruntur fellowship LQ100102201. H.R. discloses support for this work from the Max Planck Society Dioscuri Program No. LV23025.
%\bibliographystyle{naturemag}  % ama, nar, alpha, plain, chicago, abbrv, siam
%\bibliography{Refs}

\onecolumngrid

\newpage

\begin{figure}[h!]
%	\centering
	\includegraphics[width=1.0\linewidth]{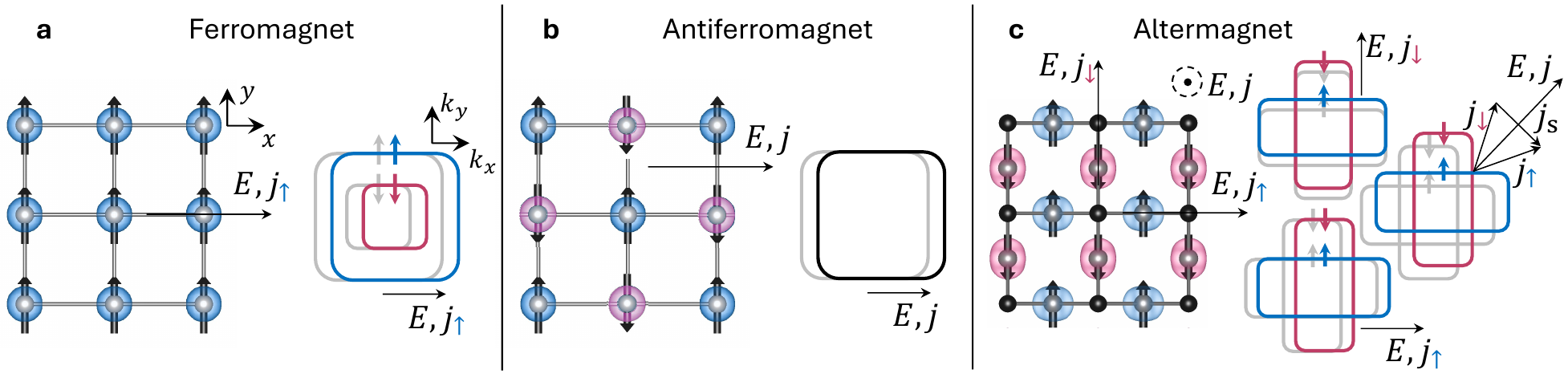}
	\caption{{\bf Comparison of altermagnets with conventional ferromagnets and antiferromagnets.} {\bf a,} Cartoons of the position-space crystal structure (left) and the momentum-space Fermi-surfaces (right) showing the generation of the spin-polarized current in a model ferromagnet. Parallel arrows and the same (blue) color of atoms represent the ferromagnetic order on the crystal. Larger/smaller grey Fermi surfaces represent majority/minority spin electronic structure in equilibrium. Blue/red spin-up/spin-down Fermi surfaces are shifted along the direction of the applied bias $E$ (horizontal black arrows). The resulting electrical current is spin-up polarized ($j_{\uparrow}$) {\bf b,} Same for a model antiferromagnet with opposite-spin sublattices (blue/red and up/down arrows) related by translation, rendering the electrical current spin unpolarized ($j$). {\bf c,} Model altermagnet with the longitudinal current for the applied bias along the $x$-axis dominated by  electrons form the spin-up sublattice ($j_{\uparrow}$), and for the bias along the $y$-axis by electrons from the spin-down sublattice ($j_{\downarrow}$). Corresponding spin-polarizations of the longitudinal currents are also shown in the momentum space\cite{Gonzalez-Hernandez2021}. For the bias applied along the in-plane diagonal, spin-up ($j_{\uparrow}$) and spin-down ($j_{\downarrow}$) electrical currents are deflected by an opposite angle form the bias direction. In the corresponding spin-splitter effect\cite{Gonzalez-Hernandez2021}, the longitudinal current is spin unpolarized ($j$) while a pure spin-current ($j_s$) flows in the transverse direction. For the bias applied in the out-of-plane direction, the current is spin unpolarized ($j$).
}
\label{fig1}
\end{figure}

%\newpage

\begin{figure}[h!]
%	\centering
	\includegraphics[width=1.0\linewidth]{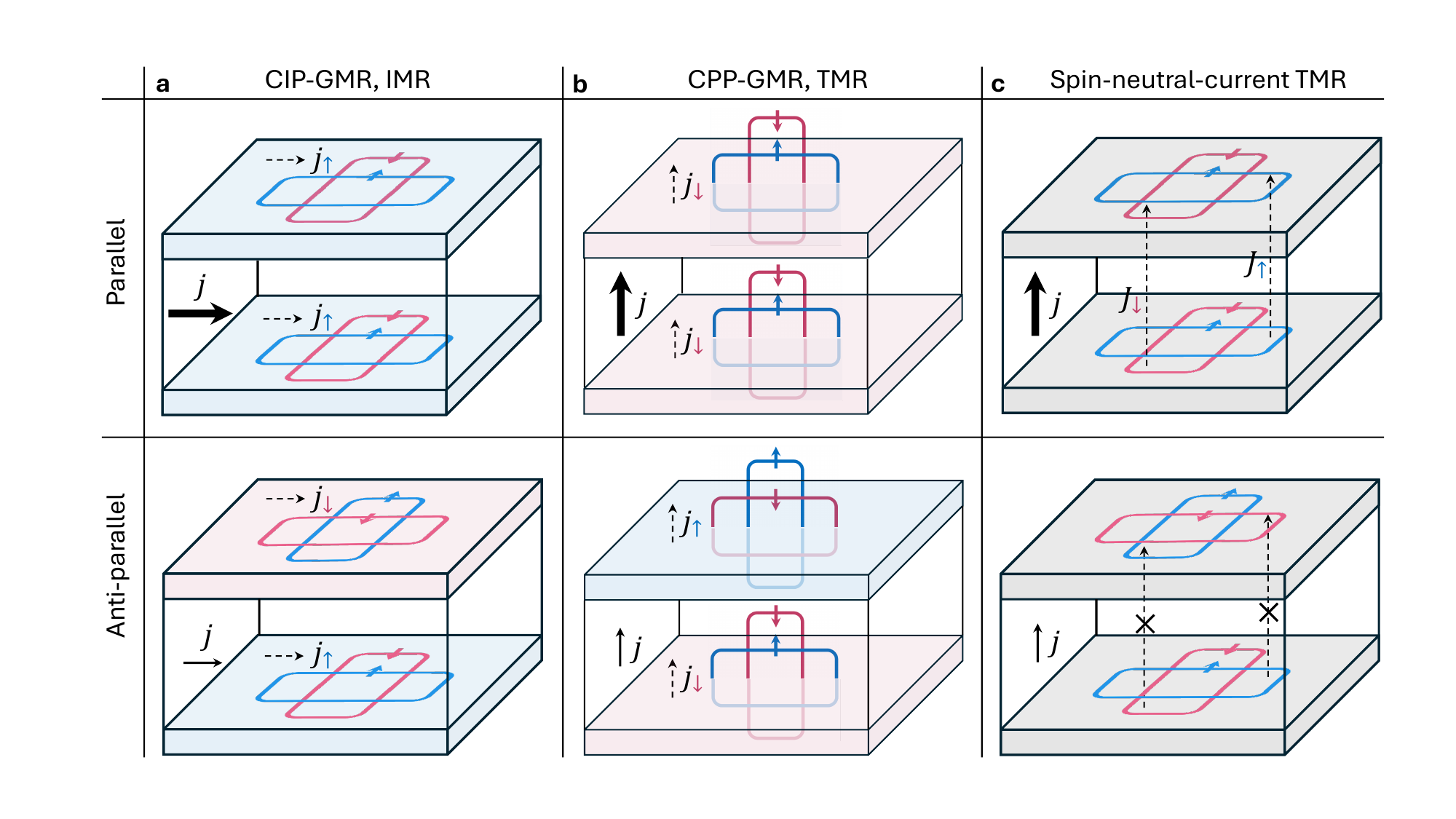}
	\caption{{\bf Altermagnetic magnetoresistance effects.} {\bf a,} Schematics of the altermagnetic counterparts of the giant magnetoresistance effect in the current-in-plane geometry (CIP-GMR)\cite{Smejkal2022GMR} in a stack comprising metallic altermagnetic electrodes separated by a metallic spacer (cf. middle panel in Fig.~3a). The altermagnetic infinite magnetoresistance (IMR) effect\cite{Giil2024} can be realized when the spacer layer is formed by a conventional superconductor (cf. top panel in Fig.~3a).  {\bf b,} Altermagnetic counterparts of the current-perpendicular-to-plane (CPP) GMR and the tunneling magnetoresistance TMR effect\cite{Smejkal2022GMR} with metallic and insulating spacers, respectively (cf. middle and bottom panels in Fig.~3a). {\bf c,} TMR in the spin-neutral-current geometry\cite{Smejkal2022GMR,Shao2021}. The top/bottom row  in a-c corresponds to the parallel/anti-parallel configuration of the altermagnetic orders in the two electrodes. Blue/red shading and corresponding $j_{\uparrow}$/$j_{\downarrow}$ labels in a and b indicate the spin-up/spin-down polarization of the electrical current in the given altermagnetic electrode. Crosses in the bottom panel of c indicate that tunneling in both spin channels is blocked in the antiparallel configuration, while in the parallel configuration tunneling in both spin channels is allowed. In a-c, large/small total current ($j$) in the parallel/anti-parallel configuration (at a fixed applied bias) is indicated by thick/thin arrow. 
}
\label{fig2}
\end{figure}

\newpage

\begin{figure}[h!]
%	\centering
	\includegraphics[width=.9\linewidth]{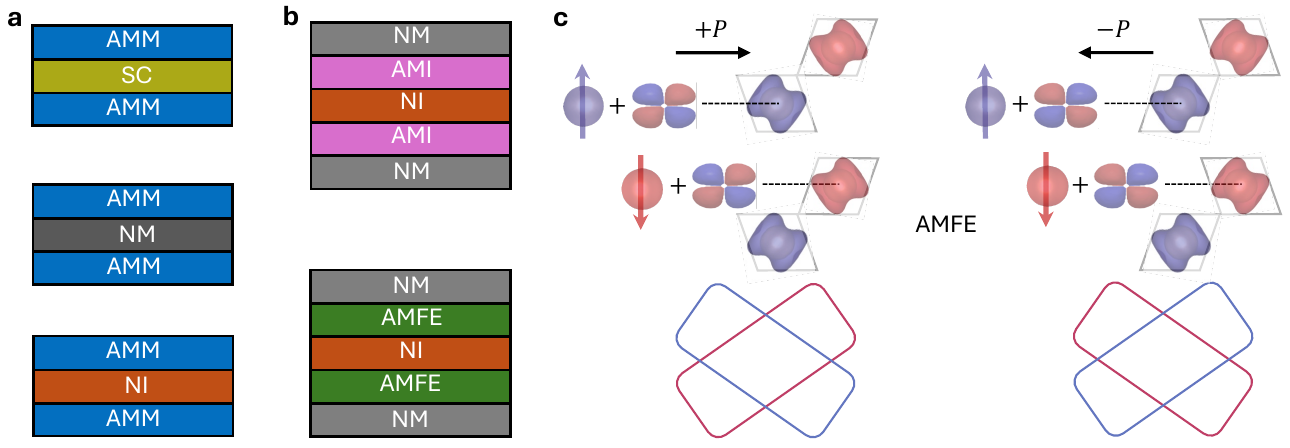}
	\caption{{\bf Altermagnetic multilayer structures and altermagneto-electric effect.} {\bf a,} Trilayer stacks corresponding to the IMR, GMR, and TMR effects (cf. Fig.~2)\cite{Smejkal2022GMR,Shao2021,Giil2024}. AMM stands for altermagnetic metal, SC for superconductor, NM for normal metal, and NI for normal insulator. {\bf b,} TMR stack with altermagnetic-insulating (AMI) layers\cite{Samanta2025} (top), and with altermagnetic-ferroelectric (AMFE) layers (bottom). Only representative examples of multilayers are shown in a and b; other multilayers can be also considered, e.g., combining different types of altermagnets in the stack.  {\bf c,} Schematics of the altermagneto-electric effect providing a strong non-relativistic coupling, mediated by lattice distortions, between the ferroelectric polarization ($P$) and the alternating order\cite{Smejkal2024}. Top part shows the position-space spin-up (blue) and spin-down (red) local electronic densities on individual atoms, decomposed into an isotropic s-wave  magnetic-dipole component (blue or red sphere with arrow), and an anisotropic $d$-wave spin-density component (blue-red shapes). When switching the sign of the ferroelectric polarization, the $d$-wave spin density component also switches sign while the dipole component remains unchanged. Bottom part shows schematically corresponding spin-up (blue) and spin-down (red) energy iso-surfaces. Their spin-polarization also reverses when reversing the ferroelectric polarization.
}
\label{fig3}
\end{figure}

%\newpage

\begin{figure}[h!]
%	\centering
	\includegraphics[width=.6\linewidth]{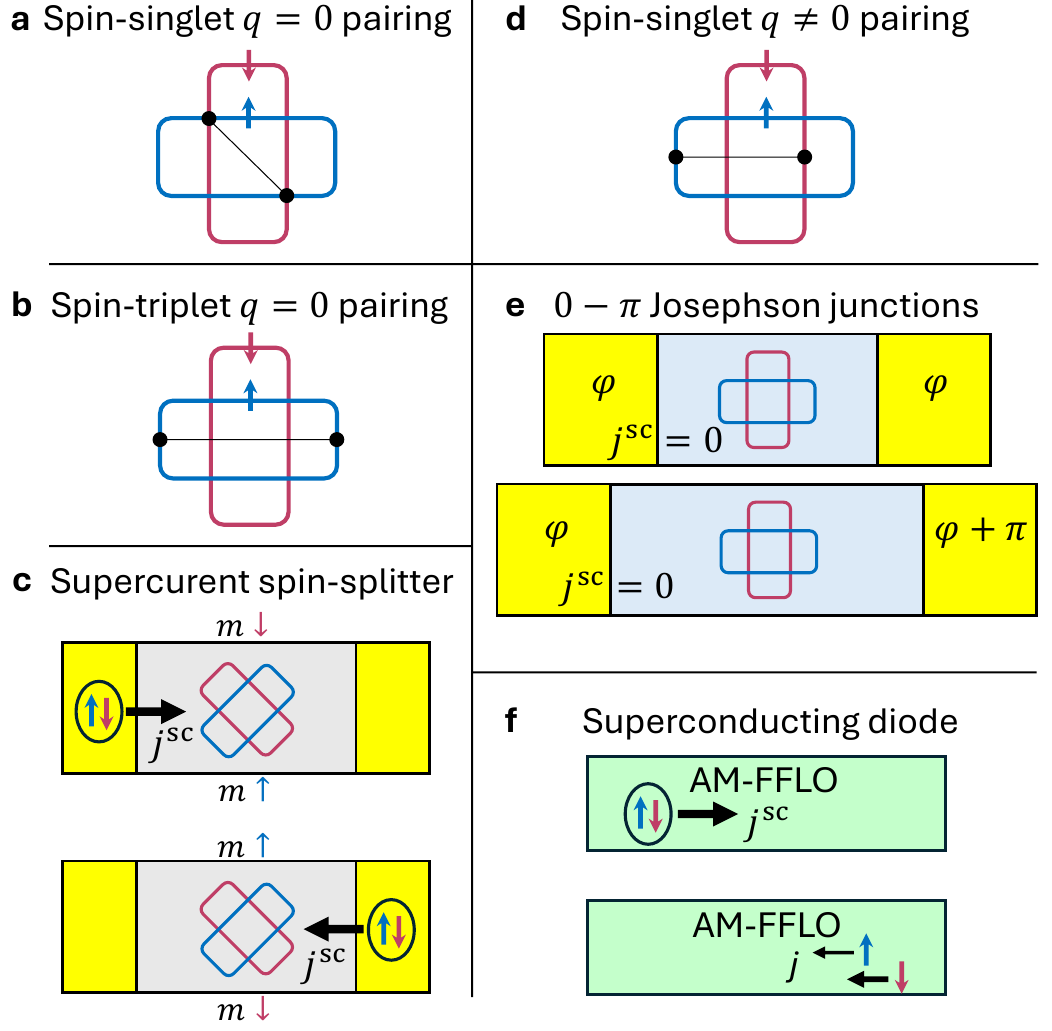}
	\caption{{\bf Interplay of altermagnetism and superconductivity.} {\bf a,b} Comparison of spin-singlet  and spin-triplet Cooper pairing  with zero center-of-mass momentum ($q=0$) in altermagnetic spin-up (blue)  and spin-down (red) Fermi surfaces. {\bf c,} Schematics of the supercurrent spin-splitter effect\cite{Giil2024a} due to the spin-triplet pairing correlations in the altermagnet (grey) separating two conventional superconducting electrodes (yellow shading). Opposite magnetizations ($m$ with opposite arrows) are generated at opposite edges of the altermagnet upon injecting the supercurrent ($j^{\rm sc}$). {\bf d,} Finite-momentum ($q\neq 0$) spin-singlet pairing. {\bf e,f} Corresponding $0-\pi$ Josephson-junction\cite{Ouassou2023,Zhang2023,Beenakker2023}  and superconducting-diode\cite{Banerjee2024a,Chakraborty2025,Yang2025,Sim2025} quantum devices. In the Josephson junctions, the ground state with zero supercurrent ($j^{\rm sc}=0$) corresponds to the same phase ($\varphi$) of the order parameters in the two superconducting electrodes, or to a phase difference of $\pi$, depending on the length of the altermagnet (blue shading) separating the two superconductors (yellow shading). In the superconducting diode, a supercurrent $j^{\rm sc}$) flows in one direction and a normal dissipative current ($j$) in the opposite direction. AM-FFLO stands for a coexisting altermagnetic phase with a superconducting Fulde-Ferrell-Larkin-Ovchinnikov phase. Encircled blue and red arrows in the top panel indicate Cooper pairs carrying the supercurrent (thick horizontal arrow). In the bottom panel with the normal current, the Cooper pairs are broken and the current is carried by dissipative spin-up and spin-down  transport channels (thin horizontal arrows).
}
\label{fig4}
\end{figure}

\end{document}